\documentclass[12pt]{iopart}
\usepackage{graphicx,graphics}

\newcommand{\nbin}    {\ensuremath{N_{\rb{bin}}}}
\newcommand{\rb}[1]   {\mbox{\textrm{\scriptsize #1}}}

\newcommand{\dndpt}   {\ensuremath{\textrm{d}N/\textrm{d}p_{\rb{T}}}}
\newcommand{\mkdnr} {\make@display@tag}
\begin{document}
\title[]{High $p_{\mathrm T}$ suppression of $\Lambda$ and $K^{0}_{s}$ in Pb--Pb collisions at $\sqrt{s_{\mathrm{NN}}}=$~2.76~TeV with ALICE}

\author{Simone Schuchmann$^1$ (for the ALICE collaboration)}

\address{$^1$ Institut f\"ur Kernphysik, Goethe-Universit\"at Frankfurt, Max-von-Laue-Str.~1, 
60438~Frankfurt am Main, Germany.}
\ead{sschuchm@ikf.uni-frankfurt.de}

\begin{abstract}
The nuclear modification factors $R_{\mathrm{AA}}$ and $R_{\mathrm{CP}}$ of $\Lambda$ and $K^{0}_{s}$ in Pb--Pb collisions at $\sqrt{s_{\mathrm{NN}}}=$~2.76~TeV measured by ALICE at LHC are presented. In central collisions a strong suppression at high $p_{\mathrm T}$ ($p_{\mathrm T}>$~8~GeV/$c$) with respect to $pp$ collisions is observed. The $p_{\mathrm T}$ region below is dominated by an enhancement of $\Lambda$ over the suppressed $K^{0}_{s}$. The results are compared to those for charged hadrons and to $\Lambda$ from lower collision energies.
\end{abstract}

\noindent{\bfseries Motivation}
\\The suppression of high transverse momentum ($p_{\mathrm T}$) hadrons in AA collisions relative to $pp$ is considered to occur due to parton energy loss in the medium before hadronization. Hence, the measurement of the amount of the suppression is an important diagnostic means to probe the medium. Nuclear modification of particle production is commonly expressed in terms of the nuclear modification factors $R_{\mathrm{AA}}$ and $R_{\mathrm{CP}}$ \cite{MLeuwen}:\\

\begin{equation*}
R_{\rb{AA}} = \frac{1}
            {\left\langle \nbin \right\rangle}
       \cdot  
       \frac{\left(\dndpt\right)_{\rb{AA}}}
            {\left(\dndpt\right)_{\rb{pp}}}
\end{equation*}
\begin{equation*}
 R_{\rb{CP}} = \frac{\left\langle \nbin^{\rb{peripheral}} \right\rangle}
            {\left\langle \nbin^{\rb{central}} \right\rangle}
       \cdot  
       \frac{\left(\dndpt\right)_{\rb{central}}}
            {\left(\dndpt\right)_{\rb{peripheral}}},
\end{equation*}
$\langle N_{\mathrm{bin}}\rangle$: number of binary NN collisions in (central or peripheral) AA collisions.\\A strong suppression of charged hadrons in central Pb--Pb collisions relative to $pp$ at $\sqrt{s_{\mathrm{NN}}}=$~2.76~TeV in the intermediate $p_{\mathrm T}$ region ($p_{\mathrm T}=$~6~--~8~GeV/$c$) followed by a rise up to $p_{\mathrm T}=$~20~GeV/$c$ was reported by ALICE \cite{Aamodt:2010jd}. The question addressed in this contribution is how the spectra of individual particle species are modified. Baryons and mesons may show a different modification pattern due to enhanced baryon production at intermediate $p_{\mathrm T}$ which is called the baryon-to-meson anomaly in the following. Identified baryon and meson particle spectra at high $p_{\mathrm T}$ may also allow to disentangle differences between the energy loss of quarks and gluons. The $\Lambda$ baryon and the $K^{0}_{s}$ meson are therefore good candidates to give a more detailed view of the parton energy loss at LHC energies. 
\newpage
\noindent{\bfseries Analysis}
\\The $\Lambda$ and $K^{0}_{s}$ are reconstructed employing a topological secondary vertex finder using tracking information  from the TPC (Time Projection Chamber) and the ITS (Inner Tracking System). For each $p_{\mathrm T}$ bin the yields are obtained by integrating over the mass peak after the combinatorial background subtraction. The latter is based on a fit of the background with a polynomial of first and second order excluding the mass peak region.\\Systematic uncertainties related to the reconstruction efficiency cancel partially in the ratios $R_{\mathrm{AA}}$ and $R_{\mathrm{CP}}$. The systematic errors indicated on the results presented below are based on a conservative estimate of the uncertainty on the centrality dependence of the reconstruction efficiency. The $p_{\mathrm T}$ spectra of $\Lambda$ are corrected for feed-down using a preliminary estimate of the contribution from $\Xi$. The uncertainty related to the centrality dependence of this estimate contributes mainly at low $p_{\mathrm T}$ and is included in the systematic errors of $R_{\mathrm{AA}}$ and $R_{\mathrm{CP}}$. \\
The following results are extracted from two centrality classes in Pb--Pb collisions\hspace{1.6cm}(0~--~5\%,~60~--~80\%) at $\sqrt{s_{\mathrm{NN}}}=$~2.76~TeV and from the corresponding $pp$ reference data at $\sqrt{s}=$~2.76~TeV.\\

\begin{figure}[t]
\centering
\begin{minipage}[b]{0.495\linewidth}
\begin{center}
\includegraphics[width=\linewidth]{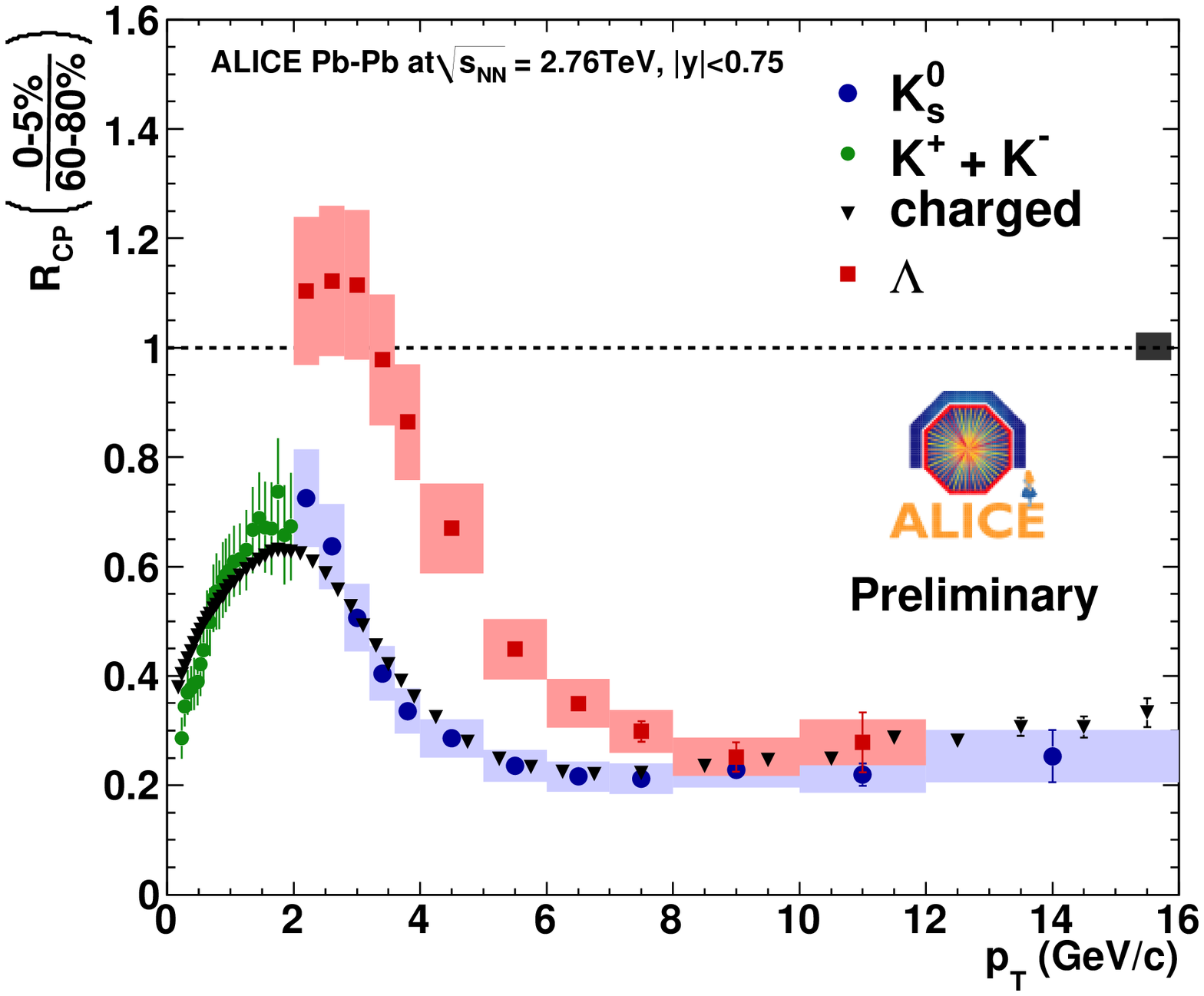}
\end{center}
\end{minipage}
\begin{minipage}[b]{0.495\linewidth}
\begin{center}
\includegraphics[width=\linewidth]{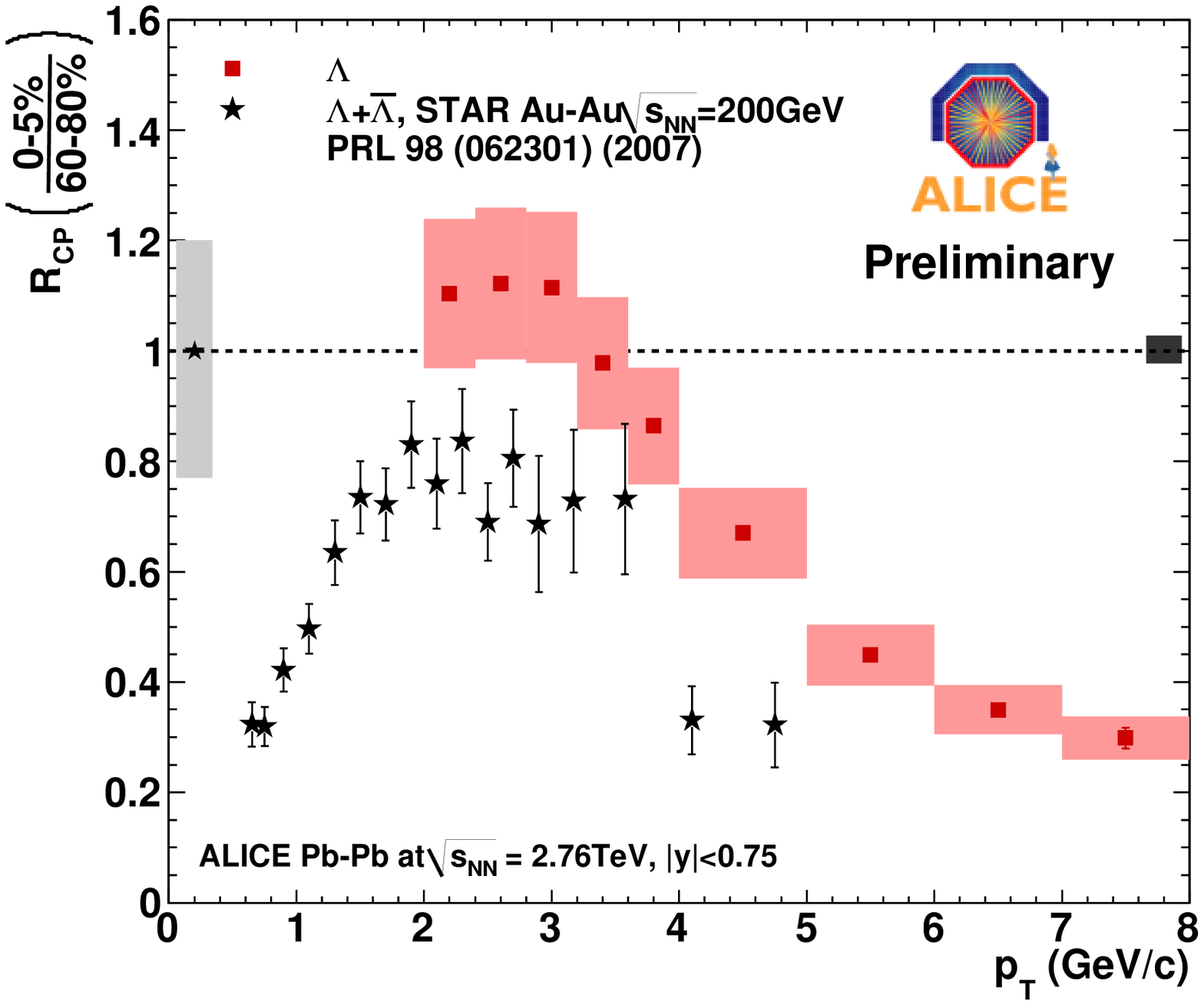}
\end{center}
\end{minipage}
\caption{The nuclear modification factor $R_{\mathrm{CP}}$ of $\Lambda$ and $K^{0}_{s}$ between
  central (0~--~5~\%) and peripheral (60~--~80~\%) Pb--Pb collisions at $\sqrt{s_{\mathrm{NN}}}=$~2.76~TeV. The boxes around the data points indicate the systematic error for $K^{0}_{s}$ and $\Lambda$. The uncertainty due to the calculation of $\left\langle \nbin \right\rangle$ is given by the gray boxes at the dotted line. Left panel: For the charged kaons (small filled circles) and the charged hadrons (filled triangles) only statistical errors are shown. Right panel: the comparison to the measurement for $\Lambda+\overline{\Lambda}$ in Au--Au collisions at $\sqrt{s_{\mathrm{NN}}}=$~200~GeV measured by the STAR collaboration \cite{Adams:2006ke}. The light gray box at unity indicates the error on $\left\langle \nbin \right\rangle$ for the STAR mesurements.}
\label{fig:rcpstar}
\end{figure}
\begin{figure}[hbt]
\centering
\begin{minipage}[b]{0.495\linewidth}
\begin{center}
\includegraphics[width=\linewidth]{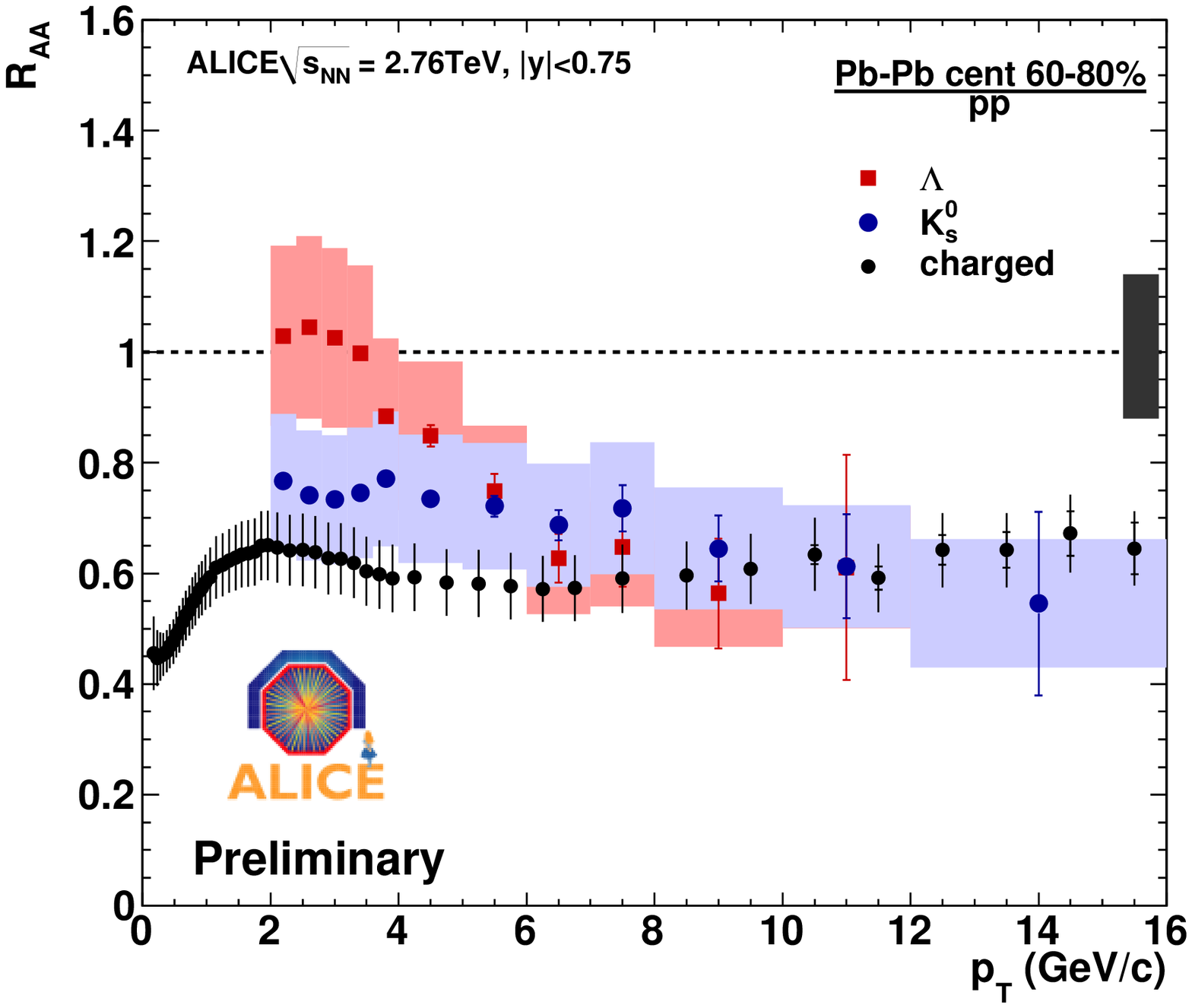}
\end{center}
\end{minipage}
\begin{minipage}[b]{0.495\linewidth}
\begin{center}
\includegraphics[width=\linewidth]{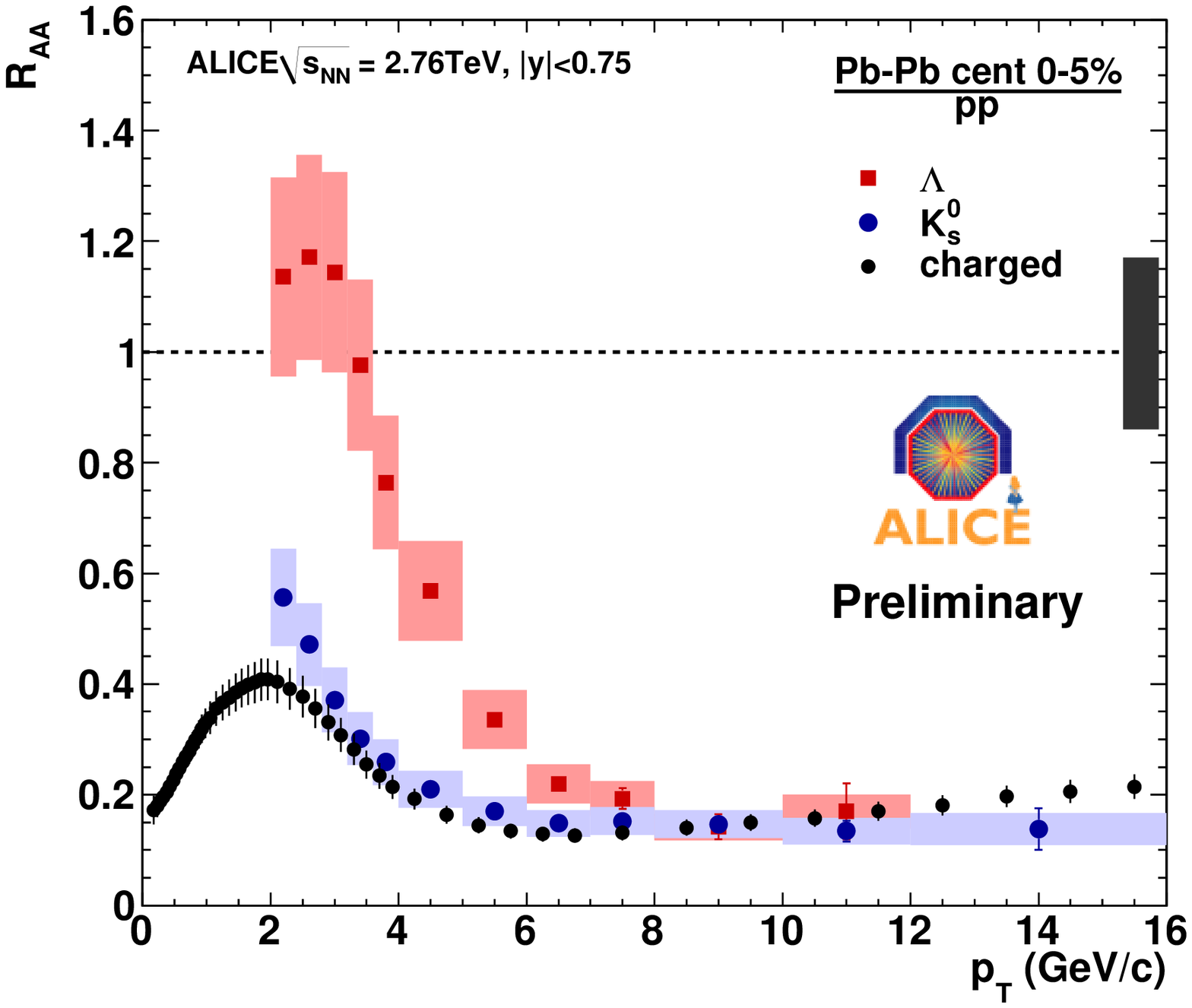}
\end{center}
\end{minipage}
\caption{The nuclear modification factor $R_{\mathrm{AA}}$ of $K^{0}_{s}$ and $\Lambda$ for peripheral (60~--~80~\%) Pb--Pb collisions (left panel) and for central (0~--~5~\%) Pb--Pb collisions at $\sqrt{s_{\mathrm{NN}}}=$~2.76~TeV (right panel). Errors as in figure \ref{fig:rcpstar}. In both panels the charged hadron $R_{\mathrm{AA}}$ is shown in addition. Vertical error bars indicate the systematic uncertainties and the horizontal ticks show the statistical errors.}
\label{fig:raaperiph}
\end{figure}
\begin{figure}[hbt]
\centering
\begin{minipage}[b]{0.495\linewidth}
\begin{center}
\includegraphics[width=\linewidth]{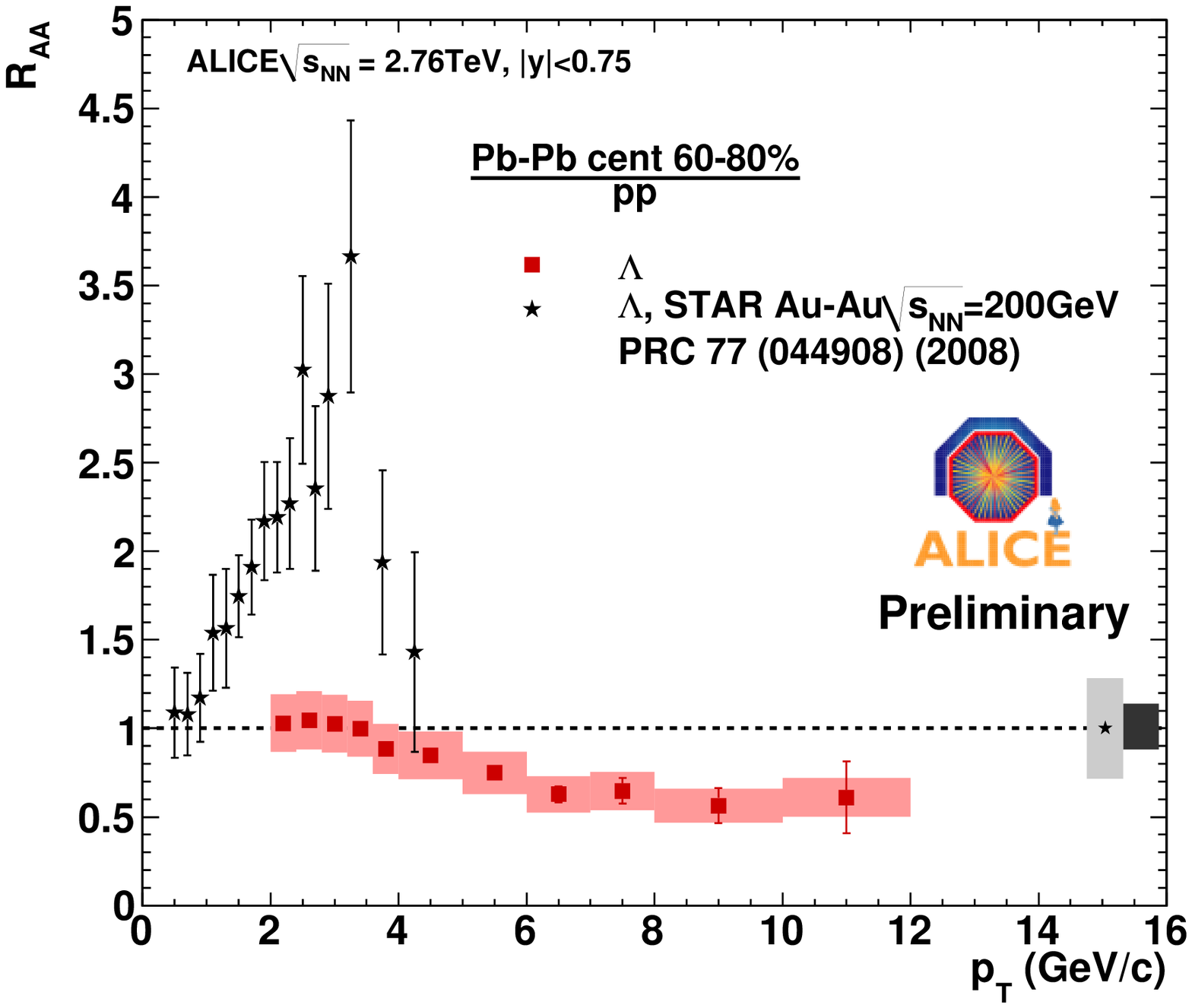}
\end{center}
\end{minipage}
\begin{minipage}[b]{0.495\linewidth}
\begin{center}
\includegraphics[width=\linewidth]{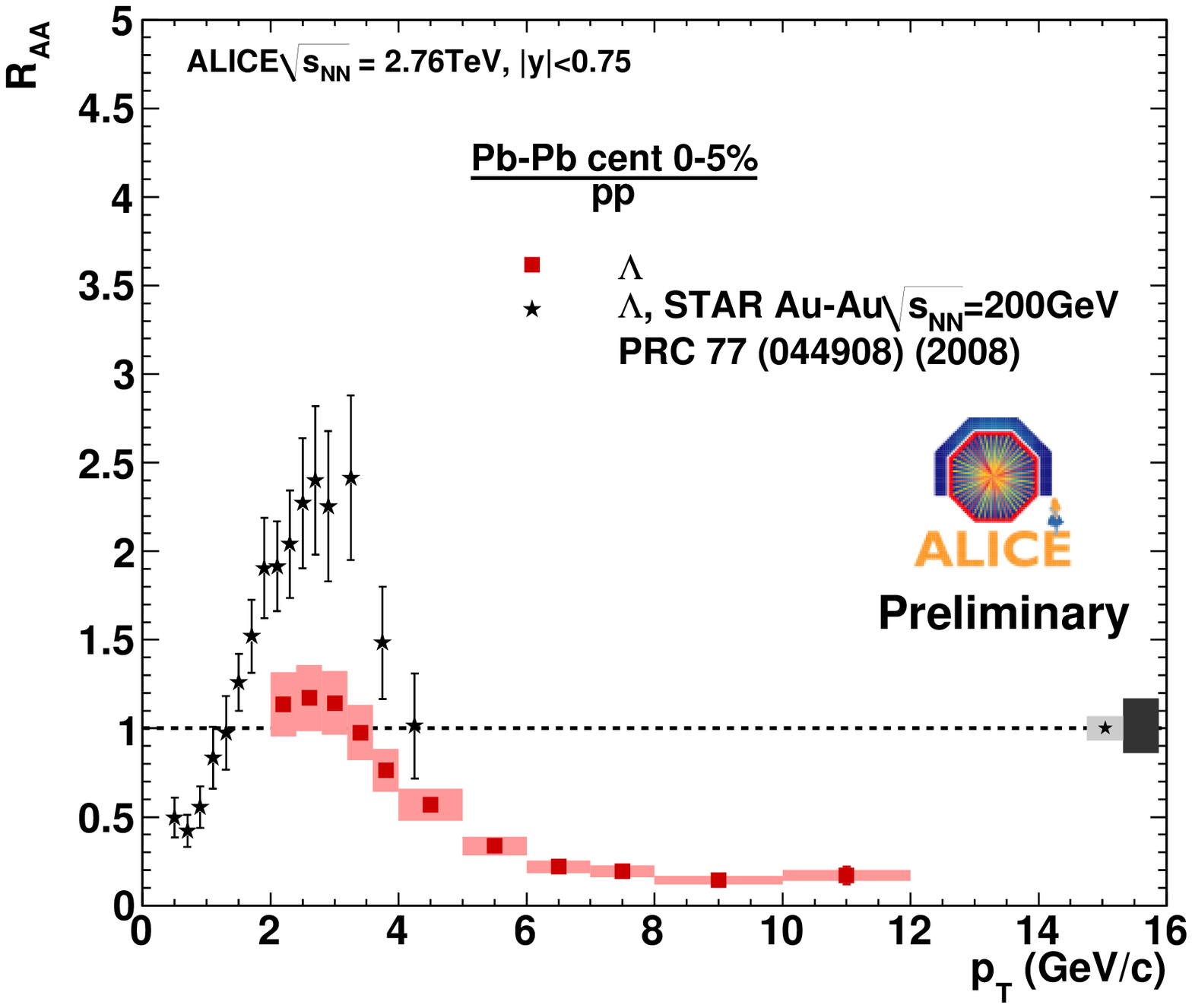}
\end{center}
\end{minipage}
\caption{The nuclear modification factor $R_{\mathrm{AA}}$ of $\Lambda$ in comparison to the measurement for Au--Au collisions at  $\sqrt{s_{\mathrm{NN}}}=$~200~GeV by the STAR collaboration \cite{Abelev:2007xp}. The left panel shows the result for peripheral (60~--~80~\%) and the right panel $R_{\mathrm{AA}}$ for central (0~--~5~\%) collisions. Errors as in figure \ref{fig:rcpstar}.}
\label{fig:raastarperiph}
\end{figure}

\noindent{\bfseries Results}
\\The left panel in figure~\ref{fig:rcpstar} shows $R_{\mathrm{CP}}$ for $\Lambda$ and $K^{0}_{s}$ compared to the $R_{\mathrm{CP}}$ of charged kaons and of charged hadrons \cite{JacekRCP}. The charged kaons match the $K^{0}_{s}$ data well at low $p_{\mathrm T}$. The $R_{\mathrm{CP}}$ of kaons is similar to the $R_{\mathrm{CP}}$ of charged hadrons over the whole $p_{\mathrm T}$ range. In particular, $K^{0}_{s}$ at high $p_{\mathrm T}$ are as much suppressed as charged hadrons. The same is true for the $\Lambda$ where at high $p_{\mathrm T}$ ($p_{\mathrm T}>$~8~GeV/$c$) a common suppression relative to $pp$ with the charged hadrons and the $K^{0}_{s}$ is observed. The enhancement of $\Lambda$- over $K^{0}_{s}$-$R_{\mathrm{CP}}$ at low-to-intermediate $p_{\mathrm T}$ might be related to the so-called baryon-to-meson anomaly observed at RHIC which is discussed in detail in \cite{JouriLK}. \\A comparison to the results obtained by the STAR collaboration for $\Lambda+\overline{\Lambda}$ at $\sqrt{s_{\mathrm{NN}}}=$~200~GeV~\cite{Adams:2006ke} is shown in the right panel of figure~\ref{fig:rcpstar}. The ALICE results are slightly larger than the STAR values and the $\Lambda$-$R_{\mathrm{CP}}$ enhancement is extended towards higher $p_{\mathrm T}$.\\In figure \ref{fig:raaperiph} the $R_{\mathrm{AA}}$ of $\Lambda$ and $K^{0}_{s}$ are shown together with the $R_{\mathrm{AA}}$ of charged hadrons \cite{JacekRCP} in peripheral (left) and central collisions (right), respectively. In peripheral events $R_{\mathrm{AA}}$ is nearly constant for $K^{0}_{s}$ and indicates a rather moderate but significant suppression of $R_{\mathrm{AA}}\approx$~0.6. Above $p_{\mathrm T}=$~7~GeV/$c$ the suppression is compatible with that of charged hadrons. For the $\Lambda$, there is little nuclear modification, $R_{\mathrm{AA}}\approx$~1, observed at\hspace{1.6cm}$p_{\mathrm T}=$~2~--~5~GeV/$c$. At higher $p_{\mathrm T}$ the suppression is similar to $K^{0}_{s}$ and to the charged hadrons. Due to this moderate $p_{\mathrm T}$ dependence of the nuclear modification factors for peripheral collisions, $R_{\mathrm{AA}}$ is expected to behave similarly to $R_{\mathrm{CP}}$ for central collisions, only being scaled by some factor. This can be seen in the right panel in figure~\ref{fig:raaperiph} where the $\Lambda$- and $K^{0}_{s}$-$R_{\mathrm{AA}}$ for central events is depicted. A strong suppression at high $p_{\mathrm T}$ with respect to $pp$ collisions is observed for both hadrons which again is similar to the suppression of charged hadrons. In the low-to-intermediate $p_{\mathrm T}$ region we again observe an enhancement of $\Lambda$-$R_{\mathrm{AA}}$ over the $K^{0}_{s}$-$R_{\mathrm{AA}}$ as it is indicated by the $R_{\mathrm{CP}}$ results.\\
Figure~\ref{fig:raastarperiph} shows the comparison of $\Lambda$-$R_{\mathrm{AA}}$ to the measurements by the STAR collaboration at $\sqrt{s_{\mathrm{NN}}}=$~200~GeV for both centralities \cite{Abelev:2007xp}. At LHC we observe a much smaller modification than at RHIC. Taking into account that the STAR $\Lambda$-$R_{\mathrm{CP}}$ is compatible to our measurement the significant difference between STAR and ALICE $\Lambda$-$R_{\mathrm{AA}}$ at $p_{\mathrm T}\approx$~3~GeV/$c$ may be driven by the $pp$ references rather than by nuclear effects. As the STAR results are limited to the intermediate $p_{\mathrm T}$ region, no statement on high  $p_{\mathrm T}$ suppression of $\Lambda$ and $K^{0}_{s}$ with respect to $pp$ collisions from RHIC to LHC energies is possible yet.\\

\noindent{\bfseries Summary}
\\We have presented the first measurements of the nuclear modification factors of $\Lambda$ and $K^{0}_{s}$ up to $p_{\mathrm T}=$~16~GeV/$c$ in Pb--Pb collisions at $\sqrt{s_{\mathrm{NN}}}=$~2.76~TeV. For both particle species a strong suppression at high $p_{\mathrm T}$ ($p_{\mathrm T}>$~8~GeV/$c$) in central collisions with respect to $pp$ collisions is found. A significant high $p_{\mathrm T}$ suppression of both hadrons is also observed in the ratio of central-to-peripheral collisions. The nuclear modification of $\Lambda$ and $K^{0}_{s}$ is compatible with the modification of charged hadrons at high $p_{\mathrm T}$. At lower $p_{\mathrm T}$ ($p_{\mathrm T}<$~5~GeV/$c$) we observe an enhancement of the $\Lambda$-$R_{\mathrm{AA}}$ with respect to the $K^{0}_{s}$-$R_{\mathrm{AA}}$ and thus to the $R_{\mathrm{AA}}$ of charged hadrons, which might be related to the baryon-to-meson anomaly observed at RHIC. While the $R_{\mathrm{CP}}$ is similar at RHIC and LHC, we find significantly less $\Lambda$-$R_{\mathrm{AA}}$ enhancement relative to $K^{0}_{s}$-$R_{\mathrm{AA}}$ at intermediate $p_{\mathrm T}$ in central and peripheral events as compared to the STAR results.
\label{startsample}

\end{document}